\documentclass[twocolumn,12pt,trackchanges]{aastex62}
\usepackage{graphics}
\usepackage{grffile}
\usepackage{rotating}

\usepackage{natbib}
\usepackage{url}

\usepackage{amsmath}  
\usepackage{amssymb,amsmath}  
\usepackage{color}
\usepackage{times}
\usepackage[normalem]{ulem}             
\usepackage{cancel}
\usepackage{verbatim}                           
\usepackage{mathrsfs}			    	
\usepackage[mathscr]{euscript}
\usepackage{fancyref}
\usepackage{hyperref}
\usepackage{physics}

\usepackage{General}
\usepackage{EM}
\usepackage{PhaseScreens}
\usepackage{ISM}

\newcommand{\thetar}{\theta_{\rm r}}
\newcommand{\td}{t_{\rm d}}
\newcommand{\tg}{t_{\rm g}}
\newcommand{\taud}{\tau_{\rm d}}
\newcommand{\dnud}{\Delta\nu_{\rm d}}
\newcommand{\nuGHz}{\nu}
\newcommand{\nuf}{\nu_{\rm f}}
\newcommand{\kDM}{C_{\DM}}
\newcommand{\kRM}{C_{\RM}}
\newcommand{\kEM}{C_{\EM}}
\newcommand{\lperp}{\textit{l}_{\perp}}
\newcommand{\lperpau}{\textit{l}_{\perp, \rm AU}}
\newcommand{\DMhat}{{\widehat{\DM}}}
\newcommand{\RMhat}{{\widehat{\RM}}}

\begin{document}

\title{{\large\bf
Induced Polarization from Birefringent Pulse Splitting in Magnetoionic Media}
\\ {\rm  \today}
}
\shorttitle{Pulse Splitting}
\shortauthors{Suresh and Cordes}

\correspondingauthor{A.~Suresh}

\email{\tt as3655@cornell.edu}

\author{A. Suresh}
\affil{Cornell Center for Astrophysics and Planetary Science and Department of Astronomy, Cornell University, Ithaca, NY 14853, USA}
\author{J. M. Cordes} 
\affil{Cornell Center for Astrophysics and Planetary Science and Department of Astronomy, Cornell University, Ithaca, NY 14853, USA}



\begin{abstract}
Birefringence in ionized, magnetized media is usually measured as Faraday rotation of linearly polarized radiation. However, pulses propagating through regions with very large Faraday rotation measures (RMs) can split into circularly polarized components with measurable differences in arrival times $\propto \nu^{-3}\,\RM$, where $\nu$ is the radio frequency. Differential refraction from gradients in DM (dispersion measure) and RM can contribute a splitting time $\propto \vert \grad_\perp \DM \vert \vert \grad_\perp\RM \vert\,\nu^{-5}$. Regardless of whether the emitted pulse is unpolarized or linearly polarized, net circular polarization will be measured when splitting is a significant fraction of the pulse width. However, the initial polarization may be inferable from  the noise statistics of the bursts. Extreme multipath scattering that broadens pulses can mask splitting effects. We discuss particular cases such as the Galactic center magnetar, J1745$-$2900, and the repeating fast radio burst source, FRB 121102. Both lines of sight have $\vert \RM \vert \sim 10^5~\RMunits$ that yields millisecond splittings at frequencies well below $\sim 1 \text{ GHz}$. We also consider the splitting of nanosecond shot pulses in giant pulses from the Crab pulsar and the minimal effects of birefringence on precision pulsar timing. Finally, we explore the utility of two-dimensional coherent dedispersion with DM and RM as parameters.     
\end{abstract}

\keywords{stars: neutron --- stars: magnetars ---  galaxies: ISM ---Fast Radio Bursts: FRB 121102 ---  radio continuum: galaxies --- scattering}

\section{Introduction}
Radio waves propagating through the interstellar medium (ISM) display dispersive arrival times following the cold plasma dispersion relation. Traditional pulse dedispersion techniques used in pulsar data analysis typically correct for this inverse-frequency squared delay that scales linearly with the dispersion measure (DM). As most pulsars are characterized by low Faraday rotation measures (RMs), time-of-arrival (TOA) corrections introduced by birefringence effects are usually negligible.

However, recent discoveries of objects with large RMs, such as the repeating FRB 121102 \citep{Spitler+14,Michilli2018} and the Galactic center magnetar J1745$-$2900 \citep{Eatough2013}, have revealed the presence of extreme magnetoionic environments in galaxies. Magnetized plasma offers different refractive indices for the two opposite modes of circular polarization. These modes propagate with different speeds and hence, develop a propagation-induced relative time delay in such media. If this delay is long enough,  a linearly polarized or unpolarized wave emitted at a transient astrophysical source may manifest to an observer as a sequence of two pulses with opposite circular polarizations. Here, we study birefringence effects arising via different mechanisms, and evaluate their significance at different observing frequencies.

Sections~\ref{sec:arrival_time} and \ref{sec:obs_effects} discuss TOA corrections due to dispersive group velocities, differential refraction and plasma lensing. In \S~\ref{sec:FRB_split}, we investigate the frequency dependence of burst shapes and polarizations for FRB 121102 by taking into account pulse broadening and scattering. Observable birefringence effects for the Galactic center magnetar are examined in \S~\ref{sec:Galcenter_magnetar}. We find that arrival time perturbations due to differential dispersion and refraction of nanoshot pulses through filaments in the Crab nebula could possibly generate the nanosecond-duration pulses \citep{Hankins2003,Hankins2007} seen from the Crab pulsar. This is discussed in \S~\ref{sec:Crab}. Accurate dedispersion of pulses received from sources embedded in extreme magneto-active environments requires knowledge of the source RM in addition to its DM. In \S~\ref{sec:deDMRM}, we consider the plausibility of a two-dimensional coherent dedispersion technique that operates to simultaneously optimize two parameters, i.e., the DM and RM. In \S~\ref{sec:noise}, we highlight the importance of exploiting burst noise statistics for inferring initial polarization states of observed pulses. Finally, we summarize our findings and present the conclusions from our study in \S~\ref{sec:summary}.

\section{Arrival Times and Refraction for a Magnetoionic Medium}\label{sec:arrival_time}

A transverse plane wave $e^{i (\kvec\cdot \xvec - \omega t)}$ propagating along the $\epsthreehat = \epsonehat\cross\epstwohat $ direction in a right-handed coordinate system is expanded into circularly polarized components
\be
\Evec_{\pm}(\xvec, t) 
	=
        \Re \left\{ E \epspm \,  e^{i (\kvec\cdot \xvec - \omega t)} \right\}
        =
        \frac{E}{\sqrt{2}}
	\left [ 
		\genfrac{}{}{0pt}{}{\cos (kz - \omega t)}{ \mp\sin (kz-\omega t) }
	\right].
\label{eq:Evec}
\ee
using the unit vectors
$\epspm =  (\epsonehat \pm i \epstwohat)/\sqrt{2}$.

In conformance with the IEEE \citep{IEEEconv} and IAU \citep[e.g.][]{1996A&AS..117..161H} conventions \citep[see also][]{2017isra.book.....T},    the $+$ (upper sign) and $-$ (lower sign) cases correspond to right-and-left hand circular polarizations (RHCP, LHCP), respectively,  and the RHCP wave is said to have positive helicity \footnote{Note however that the positive helicity wave is deemed to be left-hand circularly polarized in the optics community 
\cite[e.g.][]{1999prop.book.....B};     \cite[see also][]{jackson_99}. }.  The Stokes parameter $V = I_{\rm R} - I_{\rm L}$ is  positive if the wave is RHCP,   where $I_{\rm R, L}$ are the intensities measured in the RHCP and LHCP channels respectively, of a dual-channel receiver.

For a non-relativistic plasma consisting of electrons with number density $\nelec$ and positive charges
(ions or positrons) with number density $\npos$ in a magnetic field $B_0$ directed along the $\epsthreehat$ axis, the index of refraction, $\nr = kc/\omega$ is 
\cite[][]{1975JPlPh..13..571I}
\be
\!\!\!\!\!
\nr^2 \simeq
1 - \left(\frac{\ompe}{ \omega}\right)^2
	\left[
	\frac{1}{1\mp\omcye/\omega}
	+ 
	\frac{\npos\me/\nelec\mpos}{1\pm\omcyp/\omega}
	\right],
\label{eq:disprel1}
\ee
where the approximate equality denotes that the Lorentz force from the wave's magnetic field is assumed  negligible.
The quantities $\omcye = e\vert B_0\vert /\me c$ and $\omcyp=e\vert B_0\vert/\mpos c$ are the  unsigned cyclotron frequencies of the electron and positive charges respectively. The electron plasma frequency is
\be
\ompe = \left(\frac{4\pi \nelec e^2}{\me} \right )^{1/2}.
\ee
The phase and group velocities are
\be
\vphase =\frac{\omega}{k} = \frac{c}{\nr},
\quad \quad
\vgroup = \frac{d\omega}{dk} = \frac{c}{\nr + \omega d\nr/d\omega}.
\label{eq:vpvg}
\ee
The dispersive propagation time of a pulse is then
\be
\td = \int_0^d \frac{ds}{\vgroup} = c^{-1} \int_0^d ds\, 
	\left(\nr + \omega d\nr/ d\omega \right).
\label{eq:disp_delay}
\ee
In the following we  drop the integration limits because we are concerned only with integrals from the source to the observer. 

Assuming paraxial optics (only small angular deviations from a direct ray path), 
the phase perturbation is the integral along the direct propagation path,
\be
\phi(\xperpvec) = k \int ds\, [\nr(\xperpvec) -1],
\label{eq:phase_pert_def}
\ee
where $\xperpvec$ is a vector that is perpendicular to our line of sight to the source. 
Gradients in the plasma parameters transverse to the line of sight refract the radiation through angles
\be
{\thetavec_r}(\xperpvec) = k^{-1} \mathbf{\nabla_{\perp}} \phi(\xperpvec).
\label{eq:thetar}
\ee
The normal modes of a magnetized medium will therefore show different arrival times
$\tg \sim \deff \thetar^2/2c$
where $\deff$ is a geometry-dependent effective distance to the source that depends on the location and depth
of the medium. 
 
\subsection{Electron-proton Plasma}
For an electron-proton plasma, we ignore the contribution from the more massive  protons, yielding
\be
\nr^2 = \left( \frac{kc}{\omega}\right)^2
	\approx 1 - \left(\frac{\ompe}{\omega} \right)^2
	      \left(\frac{1}{1\pm\omcye/\omega} \right).
\label{eq:nr2longitudinal}	      
\ee
Corrections to the plasma frequency that account for the proton mass and a finite temperature \cite[][\S5.1]{1975JPlPh..13..571I} introduce factors of $\sqrt{1+\melec / m_{\rm p}}$ and $\sqrt{1+v_{\rm T}^2/c^2}$ (where $v_{\rm T}$ is the thermal speed), respectively. These corrections are smaller than 0.03\% (for $T < 3\times10^6$~K). 

\subsubsection{Dispersive Arrival Times}

Plugging $\nr$ from \Eq \ref{eq:nr2longitudinal} in \Eq \ref{eq:disp_delay}, the three leading additions to  the vacuum propagation time $d/c$ correspond 
to terms up to second order in $(\ompe/\omega)^2$ and linear in $\omcye/\omega$. Taking into account variations in electron density and magnetic field along the line of sight, the extra delay has three terms,  
\be
\td &=& 
	\frac{e^2}{2\pi \me c} \frac{ \int ds\, \nelec}{\nu^2}
	\pm \frac{e^3}{2\pi^2 (\me c)^2} \frac{\int ds\, \nelec \Bpar}{\nu^3}
    \nonumber \\
    && \quad\quad\quad\quad\quad
	+ \frac{3 e^4}{8\pi^2 \me^2 c}	\frac{\int ds\, \nelec^2}{\nu^4}.\label{eq:extra_delay}
\ee
We define the DM, RM and the emission measure (\EM) as follows: 
\begin{align}
\DM &= \int ds\, \nelec \label{eq:DMdefinition}\\
\RM &=  \frac{e^3}{2\pi \me^2 c^4}  \int ds\, \nelec \Bpar \label{eq:RMdefinition}\\
\EM &= \int ds\, \nelec^2 \label{eq:EMdefinition}.
\end{align}
Here, the integration runs from source to observer, and a positive $\Bpar$ implies a magnetic field direction towards the observer. We note that while DM and EM are determined solely by $\nelec$ along the line of sight, RM depends  on the product of $\nelec$ and $\Bpar$. Consequently, contributions to the RM from different electrons spatially distributed along the line of sight can differ significantly from that for the DM and the EM.

Using standard units for \DM \ ($\DMunits$), \RM \ ($\RMunits$), \EM \ ($\EMunits$)  and  $\nu$ (GHz), we rewrite \Eq \ref{eq:extra_delay} as
\be 
\begin{split}
  \!\!\!\!\!\!
 \td    &=  t_{\DM} \pm  t_{\RM} + t_{\EM}
 \\
 &= 
	    \kDM \left(\frac{\DM}{\nuGHz^2} \right)
	\pm \kRM \left(\frac{\RM}{\nuGHz^3} \right)
	+   \kEM \left(\frac{\EM}{\nuGHz^4} \right)  \\
\end{split}	
\label{eq:td}
\ee
with coefficients: 
\be
\kDM &=& \frac{e^2}{2\pi \me c} 
	\times {\rm
		       \left( \frac{\rm cm}{pc}\right )
	               \left(\frac{GHz}{Hz}\right)^2
	           }
	= 4.15~{\rm ms} ,
\\
\kRM &=& \frac{c^2}{\pi} 
		\times {\rm
		       \left( \frac{cm}{pc}\right )		       
	               \left(\frac{GHz}{Hz}\right)^3
	               \left( \frac{ m}{cm}\right )^2
	           }
		= 28.6~{\rm ps}, \\
\kEM &=& \frac{3 e^4}{8\pi^2 \me^2 c} 
		\times {\rm
		       \left( \frac{\rm cm}{pc}\right )
	               \left(\frac{GHz}{Hz}\right)^4
	           }
	= 0.251 ~{\rm ps}. 
\ee

\subsubsection{Phase Perturbations from Birefringent Refraction}
To first order in $(\ompe/\omega)^2$ and $\omcye/\omega$, the phase perturbation (defined in \Eq \ref{eq:phase_pert_def}) of an electromagnetic wave propagating through a magnetized plasma is
\be
\phi(\xperpvec) &=& 
	-\lambda r_{\rm e} \int ds\, \nelec(\xperpvec)
\nonumber \\
&&\quad\quad
    \mp \frac{e^3\lambda^2}{2\pi \melec^2 c^4}
    \int ds\, \nelec(\xperpvec) \Bpar(\xperpvec),
\label{eq:phi1}
\\
&\equiv& -\lambda r_{\rm e} \DM \mp\lambda^2 \RM,
\label{eq:phi2}
\ee
 Employing standard units of $\DMunits$ and $\RMunits$ for \DM\ and \RM, respectively, and expressing frequencies in GHz, the phase is
\cite[][]{1975JPlPh..13..571I}
\be
\phi \approx 
-2.61\times 10^7 \, \nu^{-1} \DM
\mp 0.0899\, \nu^{-2} \RM \ {\rm rad}.
\label{eq:phi_dm_rm}
\ee

The standard expression for RM given in Equation~\ref{eq:RMdefinition} assumes that the normal modes of the magnetized plasma are circularly polarized and that the RHCP and LHCP  waves propagate independently  with  slightly different phase velocities given by Equations~\ref{eq:vpvg} and \ref{eq:nr2longitudinal}. Independent propagation requires that significant changes in index of refraction occur on length scales much larger than a wavelength \citep[e.g.][]{1960ApJ...131..664C}. If there are such rapid variations, \citet[][]{2010ApJ...718.1085B} have shown that Faraday rotation involves the magnitude of the total magnetic field rather than just the line of sight component.   As yet there is no indication that mode coupling occurs in any of the astrophysical sources we consider, but the possibility exists that it may occur in the extreme magnetoionic environment around the  sources of  FRB~121102 and other FRBs.

\subsection{Electron-positron and Relativistic Plasmas}
A neutral electron-positron plasma with $n_{e^+} = \nelec$ is uni-refringent (i.e. not birefringent) because the helical electron and positron currents cancel. Such a plasma is isotropic for wave propagation. The electron plasma frequency increases by a factor of $\sqrt{2}$ and the index of refraction is simply
\be
\nr^2  \approx 1 - \left(\frac{\ompe}{\omega} \right)^2
	      \left(\frac{2}{1 - \omcye^2/\omega^2} \right).
\ee
Consequently the \RM\ terms in the phase, refraction angle, and arrival time vanish. For the arrival time, the terms involving $\DM$ and $\EM$ are factors of two and four larger, respectively, 
\be
\kDM  \to  8.3~{\rm ms},
\quad
\kRM \to 0, 
\quad
\kEM \to 0.50 ~{\rm ps}. 
\ee

We note also that for a relativistic plasma, the effective masses are larger by a factor equal to the Lorentz factor, $\gamma = (1 - v^2/c^2)^{-1/2}$, so the \DM\ term will scale as $\gamma^{-1}$ while the $\RM$ and $\EM$ terms will scale as $\gamma^{-2}$. 

\section{Observable Effects}\label{sec:obs_effects}

\subsection{Differential Arrival Times}

Previous authors have  considered modifications to the inverse quadratic $t_{\DM} \propto \nu^{-2}\DM$ delay  characteristic of pulsars and FRBs.  Shortly after the discovery of pulsars, \citet{1968Sci...160..760T} constrained the EM of dispersive gas using the $\nu^{-4}$ term in \Eq\ref{eq:td}. More recently, \citet{2014MNRAS.443L..11D} made the same conclusion as \citet{1968Sci...160..760T} for the lines of sight to FRBs that dispersion was caused by a tenuous plasma. \citet{pw92} discussed the three delays in their analysis of temporal variations of \DM\  for nearby pulsars and concluded that the higher-order RM and EM  terms were negligible. Very large RM values for the repeating FRB and the Galactic center magnetar J1745-2900 clearly alter this conclusion. 
 \\
 \begin{figure}[t!]
\includegraphics[width=\linewidth]{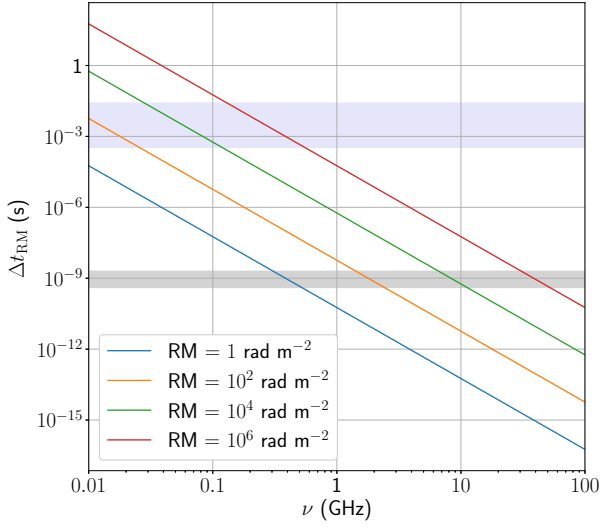}
\caption{Variation of the splitting time  $\Delta t_{\RM}$ with frequency for values of RM in the range from $1-10^6 \ \text{rad m}^{-2}$. The violet and gray horizontal bands correspond to the ranges of intrinsic temporal widths} of FRBs (\citet{Petroff2016}, \url{http://www.frbcat.org/}) and Crab nanoshots \citep{Hankins2003,Hankins2007} respectively. \label{fig:trm_vs_nu}
\end{figure}

The sign difference of the  $t_{\RM}$ term in Equation~\ref{eq:td} splits the RHCP and LHCP components of an arbitrary polarized electromagnetic wave.  Figure~\ref{fig:trm_vs_nu} shows the magnitude of the splitting time, $\Delta t_{\RM} = 2t_{\RM}$, as a function of frequency for different values of RM. At $\nu \geq 100 \ \text{MHz}$, we require $\RM \gtrsim 10^4 \ \RMunits$ to obtain values of $\Delta t_{\RM}$ that are comparable to the typical intrinsic FWHM of FRBs. The repeating FRB  121102 with measured $\RM \sim 10^5 \ \RMunits$ \citep{Michilli2018} is then a notable example where birefringence can manifest as pulse splitting at low radio frequencies. \S~\ref{sec:FRB_split} discusses the details of this pulse splitting phenomenon for FRB 121102, where pulse broadening from multipath scattering in the ISM of its host galaxy and the Milky Way is also included. We note that most FRBs possess small RMs \citep{Michilli2018} that are comparable to those of Galactic pulsars, and are hence, expected to exhibit negligible birefringent pulse splitting. However, the one-zone model for FRB 121102 developed by \citet{Margalit2018} hints at  $\RM \gtrsim 10^7 \ \RMunits$ for young FRB sources (age of $1-10$ years), suggesting potentially detectable pulse splitting effects at $\nu \lesssim 1 \ \text{GHz}$. \\ 

While Figure \ref{fig:trm_vs_nu} shows the range of intrinsic temporal widths of FRBs and Crab nanoshots, observed pulses are likely to be additionally widened due to scattering and dispersion. In particular, intrachannel DM smearing ($\sim 5$ ms for FRB 121102 at $\nu = 100 \text{ MHz}$ with a channel bandwidth of $1 \text{ kHz}$) that becomes significant at low radio frequencies can mask any visible effects of birefringence. A two-dimensional coherent dedispersion technique (refer \S~\ref{sec:deDMRM}) then serves useful to unwrap the dispersive phase delay introduced by the $t_{\DM}$ and $t_{\RM}$ terms in \Eq \ref{eq:td} and recover the dedispersed pulse profile.

\subsection{Differential Refraction}\label{sec:refraction}

\begin{figure}[t!]
\includegraphics[width=\linewidth]{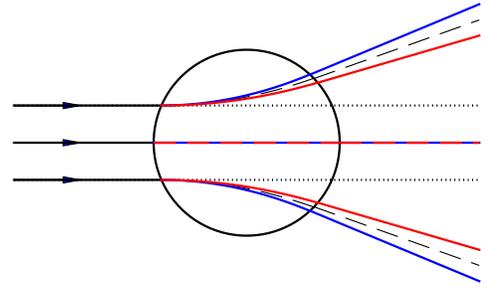}
\caption{Ray traces showing differential refraction (solid) lines for a magnetized plasma filament shown in cross section.  For $\Bpar > 0$, RHCP (blue) is refracted more than LHCP (red) and the angular difference is $\Delta{\thetar}_{,\RM}$.   The dashed lines show the refracted rays at an angle ${\thetar}_{,\DM}$ in the absence of a magnetic field. The dotted lines represent unrefracted ray paths.
\label{fig:filament}
}
\end{figure}

To illustrate birefringent refraction, we consider a filament, shown in cross section in Figure~\ref{fig:filament}, that has uniform density and a uniform magnetic field with a component along the line of sight, which is transverse to the filament axis. Gradients are non-zero only in the plane transverse to this axis. So, refraction is confined to this plane. 

Plugging Equation~\ref{eq:phi_dm_rm}
in Equation~\ref{eq:thetar} and approximating the gradients of the \DM\ and \RM\ of the filament as zero-mean quantities with typical RMS values, $\sim\DM_\text{f} / \lperp$ and $\sim \RM_\text{f} / \lperp$ respectively, and using $\lperp$ in AU, 
we obtain the refraction angles in the plane shown in Figure~\ref{fig:filament},
${\thetar}_{\pm} = {\thetar}_{,\DM} \pm {\thetar}_{,\RM}$
where
\be 
{\thetar}_{,\DM} &\approx&    
	1.715 \, (\nu^2 \lperpau)^{-1}\,\DM_\text{f} ~{\rm arc sec}
\nonumber \\
{\thetar}_{,\RM} &\approx&
     5.913\times 10^{-9} (\nu^{3}\lperpau)^{-1} \RM_\text{f} ~{\rm arc sec}.\label{eqn:thetar}
\ee

Refraction is overwhelmingly dominated  by \DM\ for extreme as well as typical values of \DM\ and \RM\ for known sources. However, the small splitting angle between the two polarizations,
\be
\Delta{\thetar}_{,\RM} \approx  1.183\times 10^{-8} 
	(\nu^{3} \lperpau)^{-1}\RM
    ~{\rm arc sec},
\ee
yields a differential time of arrival (TOA) that may be detectable. Consider a source located at a distance $\dso$ from an observer. Suppose a thin filament or lens is present at distances $\dsl$ and $\dlo = \dso-\dsl$ from the source  and the observer respectively. Assuming that this filament has dispersion measure $\DM_{\text{f}}$ and rotation measure $\RM_{\text{f}}$, the geometric TOA difference is
\be
\Delta \tg &\approx& 
	\left(\frac{\dsl\dlo}{2c\dso}\right)
	\left(
		{\thetar}_+^2 - {\thetar}_-^2 
    \right)
\\
&\approx& \left(\frac{\dsl\dlo}{c\dso}\right)
		{\thetar}_{,\DM}
        \Delta{\thetar}_{,\RM}
\\
&\approx&
49.1~{\rm ns} \, 
    \frac{{\dsl}_{,\rm kpc}\DM_{\text{f}} \ \RM_{\text{f}}}
    {\lperpau^{2} \nu^{5}}   
    \left(\frac{\dlo}{\dso} \right).   
\ee
Expressing $\DM_{\text{f}}$ and $\RM_{\text{f}}$ as fractions $f_{\DM}$ and $f_{\RM}$ of the net source DM and RM respectively, we rewrite $\Delta t_g$ as:
\be
\Delta t_g = 49.1~{\rm ns} \, 
    \frac{{\dsl}_{,\rm kpc}f_{\DM} f_{\RM}\DM \ \RM}
    {\lperpau^{2} \nu^{5}}   
    \left(\frac{\dlo}{\dso} \right).
\ee
 \begin{figure}[t!]
\includegraphics[width=\linewidth]{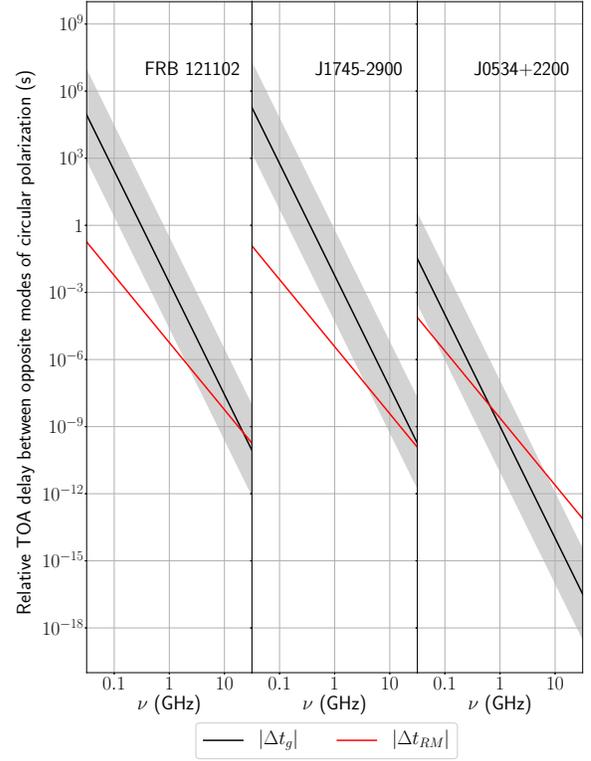}
\caption{
Spectral variation of the magnitudes of $\Delta t_{\RM}$ (red solid line) and $\Delta \tg$ (black solid line) for three systems, namely the repeating FRB 121102 (left),  the Galactic center magnetar J1745$-$2900 (middle) and the Crab pulsar J0534$+$2200 (right). In each panel, the black solid line for $|\Delta \tg |$ assumes an $\lperp$ of $10 \text{ AU}$, whereas the light grey band represents $|\Delta \tg |$ values for $\lperp \in [1 \text{ AU, }100 \text{ AU}] $. All estimates of $|\Delta \tg |$ are derived presuming $\dsl = 1 \text{ kpc}$ and $\dlo/\dso \sim 1$. For FRB 121102 and J1745$-$2900, $\Delta t_g$ is derived using $f_{\RM} = 1$ and $f_{\DM} = 0.1$. For the Crab pulsar, $\DM_{\text{f}} = 0.1 \ \DMunits$ and $\RM_{\text{f}} = -21 \ \RMunits$ are assumed .
\label{fig:f3}}
\end{figure}

Detailed modeling of the magneto-ionic environment around FRB 121102 ($\DM = 557 \ \DMunits$ \citep{Spitler+14}, $\RM \sim 10^5 \ \RMunits$ \citep{Michilli2018}) and the Galactic center magnetar ($\DM \approx 1778 \ \DMunits$, $\RM \approx -6.6 \times 10^4 \ \RMunits$ \citep{Eatough2013}) suggests that the large RMs of these sources arise from highly magnetized regions in their vicinity \citep{Margalit2018, Desvignes2018}. However, their DMs are integrated quantities with contributions from all electrons along the line of sight. To evaluate $\Delta t_g$ for these sources, we therefore assume $f_{\RM} = 1$ and $f_{\DM} \simeq 0.1$. For the Crab pulsar, we calculate $\Delta t_g$ at different frequencies presuming $\DM_{\text{f}} \simeq 0.1 \ \DMunits$ (DM depth of filaments in the Crab nebula \citep{GrahamSmith2011}) and $\RM_{\text{f}} = -21 \ \RMunits$ (average large-scale RM across the Crab nebula \citep[][]{1991ApJ...368..231B}).

Figure~\ref{fig:f3} shows the variation of $|\Delta \tg|$ and $|\Delta t_{\RM}|$ with frequency for the repeating FRB 121102, the Galactic center magnetar and the Crab pulsar. The stronger $\nu^{-5}$ dependence of $\Delta \tg$ relative to the $\nu^{-3}$ scaling of $\Delta t_{\RM}$ causes $\Delta \tg$ to become the dominant TOA splitting at frequencies below $1 \text{ GHz}$ for these three systems. 

How $\Delta\tg$ is manifested depends on the solid angle of the incident radiation (a wide or narrow beam) and details in the structure of the medium. A strong magnetoionic region may account for a (possibly small) fraction of the total  DM and RM measured for a source. Nonetheless, small scale structure on AU scales can produce measurable effects for nanoshot pulses like those seen from the Crab pulsar  \citep{Hankins2003,Hankins2007}, as shown in Figure~\ref{fig:f3} and discussed in \S~\ref{sec:Crab}.
 \begin{figure}[t!]
\includegraphics[width=\linewidth]{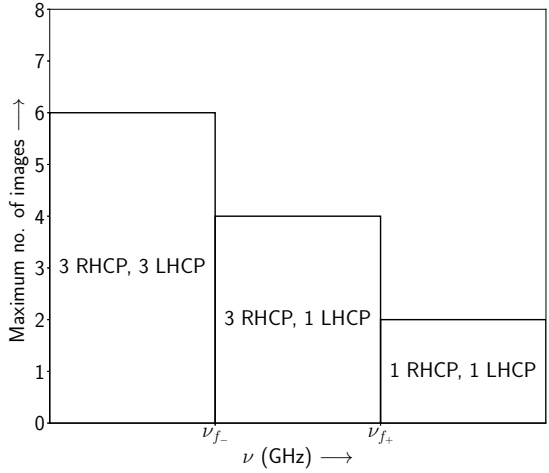}
\caption{
Different spectral regimes with varying maximum possible number of images of each circularly polarized mode. For this figure, we assume a positive $\RM_{\ell}$. Then,  $\nu_{\rm f_+}$ and $ \nu_{\rm f_-}$ correspond to the focal frequencies for RHCP and LHCP respectively. \label{fig:regimes}
}
\end{figure}

\subsection{Precision Pulsar Timing}
Millisecond pulsars used in pulsar timing arrays typically have small DMs and RMs, so polarization splitting due to differential group velocity and refraction will be small. Because splitting is symmetric, the Stokes-I pulse used for TOA estimation will be symmetrically broadened and the TOA will be minimally affected at the second order level in $\max (\Delta t_{\RM}, \Delta t_g) / W \ll 1$, where $W$ is the pulse width. 

\subsection{Plasma Lensing}
Refraction can also cause ray crossing and caustics that lead to intensity fluctuations through constructive and destructive interference.  `Extreme scattering events' (ESEs) of extragalactic sources and Galactic pulsars are attributed to  lensing structures in the galaxy \citep[][]{fdjh87, fdj+94, lrc98, 2015ApJ...808..113C,Bannister2016}. Studies of Gaussian lenses have shown features similar to those seen from ESEs \citep[e.g.][]{1998ApJ...496..253C}, and plasma lensing has been invoked as a cause for intermittency of FRBs \citep[e.g.][]{Cordes+17}.  
Recent work has identified lensing in large-amplitude pulses (magnified above the mean flux density by a factor of order $10$) from the  millisecond pulsar B1957+20 caused by evaporated gas from the pulsar's companion \citep{Main2018}.

Previous studies of plasma lensing have considered only the DM contribution to the phase and refraction angle. For such cases, a principal quantity for assessing a specific lens is its `focal frequency' $\nuf$. At frequencies lower than $\nuf$, an observer will see multiple images, and caustics will appear at and near $\nuf$. A key dimensionless parameter appearing in the lens equation of a Gaussian lens is \citep{Cordes+17}:
\be
\alpha_{\DM} = \frac{\lambda^2 r_e \DM_{\ell} }{\pi a^2}\frac{\dsl\dlo}{\dso} 
\ee
Here, $a$ is the $1/e$ radius of the lens and $\DM_{\ell}$ is the maximum DM produced by the lens. For a Gaussian lens, $\nuf$ is the frequency at which $\alpha_{\DM} = e^{3/2}/2$  yielding 
\be
\nuf  &= &	\nu \left( \frac{2\alpha_{\DM}}{e^{3/2}} \right)^{1/2} = \frac{c}{a} 
		\left( 
			\frac{2 r_{\rm e} \DM_{\ell}}{\pi e^{3/2} } 
		\right)^{1/2} \\
	& \approx&  39.1\ {\rm GHz} \times 
	\left( \frac{ \DM_{\ell}}{1 \ \DMunits} \right)
    \left( \frac{a}{1 \ \text{AU}} \right)^{-1}
	\left( \frac{\dsl\dlo/\dso}{1\ {\rm kpc}} \right) 
    \nonumber \label{eqn:focal_freq}
\ee

A magnetized plasma lens possesses a rotation measure depth $\RM_{\ell}$ in addition to a dispersion measure depth $\DM_{\ell}$. The presence of a non-zero $\RM_\ell$ can give rise to different focal frequencies for the two opposite hands of circular polarization. From Equation~\ref{eqn:thetar}, we observe that $\theta_{r,\RM}$ can be derived from $\theta_{r,\DM}$ by simply replacing $r_e \DM$ with $c\ \RM/\nu$. Performing a similar replacement of terms in $\alpha_{\DM}$, the dimensionless parameter $\alpha_{\pm}$, appearing in the lens equation for a magnetoionic lens is:
\begin{align}
\alpha_{\pm} &= \alpha_{\DM} \pm \alpha_{\RM} 
             = \alpha_{\DM} \left(1 \pm \frac{c \RM_{\ell}}{\nu r_e \DM_{\ell}}
            \right). 
\end{align}
Using standard units for $\RM_\ell$ and $\DM_\ell$, and expressing $\nu$ in GHz,
\be
\frac{\alpha_{\RM}}{\alpha_{\DM}} = 3.45 \times 10^{-9} \frac{\RM_{\ell}}{\DM_{\ell} \nu}. \label{eqn:RMvsDMalpha}
\ee
The focal frequencies $\nu_{\rm f_\pm}$ are the positive, real solutions of the cubic equations $\alpha_{\pm} = e^{3/2}/2$. We adopt a perturbative approach to solve these equations by exploiting $\alpha_{\RM} / \alpha_{\DM} \ll 1$ (Equation~\ref{eqn:RMvsDMalpha}). Denoting the focal frequency for the unmagnetized case as $\nu_{\rm f_0}$,  we have for the magnetized  lens  $\nu_{\rm f_\pm} = \nu_{\rm f_0} \pm \nu_{\rm f_1}$, where $\nu_{\rm f_1} / \nu_{\rm f_0} \ll 1$. To first order in  $\nu_{\rm f_1} / \nu_{\rm f_0}$, we obtain
\begin{align}
\nu_{\rm f_1} &= \frac{1}{2} \left( \frac{c \RM_{\ell}}{r_e \DM_{\ell}} \right) \\
& \approx 1.72 \ \text{Hz} \left( \frac{ \RM_{\ell}}{1\ \RMunits} \right) \left( \frac{ \DM_{\ell}}{1\ \DMunits} \right)^{-1} .
\end{align}

The splitting between the focal frequencies for the two opposite hands of circular polarization is $\Delta \nuf = 2\nu_{\rm f_1}$. For some of the filaments associated with the Crab pulsar J0534+2200, $\DM_{\ell} = 0.1 \ \DMunits$ \citep{Cordes+17}. Assuming $|\RM_{\ell}| \sim 21  \ \RMunits$, $\Delta \nuf \approx 0.73 \ \text{kHz}$, which is negligible in comparison to typical receiver 
channel bandwidths ($\sim 0.1 - 10 \ \text{MHz}$). However, for FRB 121102, the large $\RM_{\ell} \sim 10^5 \ \RMunits$ and an assumed $\DM_{\ell} \sim 1 \ \DMunits$ for the extreme magnetoionic environment near the source imply a $\Delta \nuf \approx 0.34 \ \text{MHz}$. Measurements that resolve the focal-frequency splitting
will show  distinct frequency regimes with differing image numbers and polarizations. \\

For  an unmagnetized lens with a  Gaussian DM profile, the lens equation gives one solution for $\nu > \nuf$, and three solutions for $\nu < \nuf$. Let us now consider a magnetized, Gaussian plasma lens. Assuming a positive $\RM_{\ell}$, Figure~\ref{fig:regimes} shows three distinct regions in the frequency spectrum which differ in the maximum possible number of images. For a given polarization, a maximum of three images is produced if the observing frequency is greater than the corresponding focal frequency. As a result, at frequencies lower than the focal frequency for LHCP radiation, $\nu_{\rm f_-}$, a maximum of six images can be obtained. On moving up the frequency axis into successive regimes, the maximum possible number of images decreases by two. At a given frequency, we see the greatest number of images (and hence, pulses) only if the relative TOA differences between them are much larger than the image pulse width. If not, a superposition of a LHCP image with a RHCP image may manifest as a single linearly polarized or unpolarized image. The exact polarization state of such an image may be inferred from the burst noise statistics. This is discussed in \S~\ref{sec:noise}. \\

Of particular interest in Figure~\ref{fig:regimes} is the spectral regime between the two different focal frequencies. For equal-amplitude circularly polarized images whose relative TOA difference is smaller than the intrinsic pulse width, we expect Stokes-V to be zero for $\nu > \nu_{\rm f_+}$ and $\nu < \nu_{\rm f_-}$. For intermediate frequencies $\nu_{\rm f_-} \le \nu \le \nu_{\rm f_+}$, the differing number of RHCP and LHCP images can lead to a non-zero Stokes-V profile, which is infeasible at other frequencies. However, we note that image production close to caustics can result in large magnifications that may differ between RHCP and LHCP modes. In addition, departures from a Gaussian lens geometry or the presence of multiple birefringent lenses can yield a significantly greater number of images of both polarizations. These factors can thus, potentially alter the spectral evolution of Stokes-V from the behavior discussed here.

\section{The Repeating FRB~121102}\label{sec:FRB_split}
\begin{figure*}[t!]
\begin{center}
\includegraphics[width=0.9\linewidth]{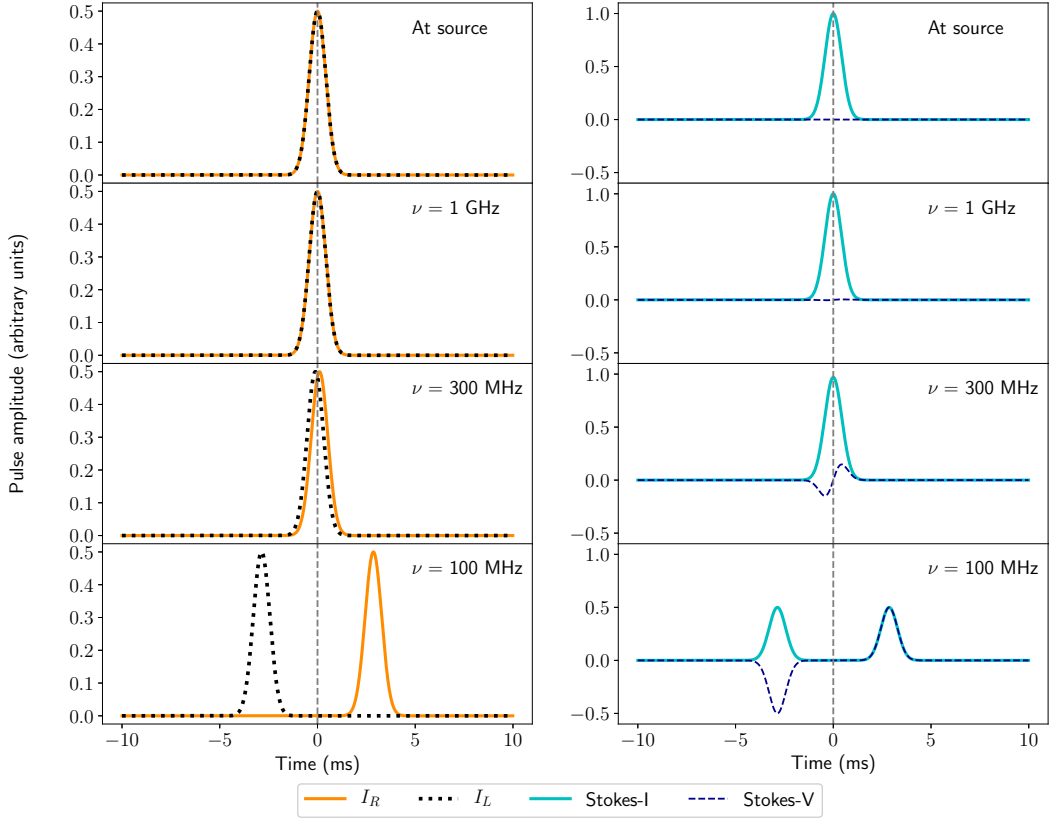}
\caption{
Left column: Model pulse profiles in $I_R$ (orange solid line) and $I_L$ (black dotted line).    
Right column: Pulse profiles in Stokes-I (cyan solid line) and Stokes-V (dark blue dashed line). 
Top row: Pulse profiles as produced at FRB source. 
Second row from top: Pulse profiles as seen by an observer at $\nu = 1 \text{ GHz}$. 
Third row from top: Observed pulse profiles at $\nu = 300 \text{ MHz}$. 
Bottom row: Observed pulse profiles at $\nu = 100 \text{ MHz}$. The vertical gray dashed line in all panels represents the time origin. Only the $\Delta t_{\RM}$ term in Equation~\ref{eq:td} has been incorporated in calculations of the pulse arrival time delays at the above frequencies. 
\label{fig:lrprofiles}}
\end{center}
\end{figure*}
Owing to the large RM and the nearly 100\% linear polarization, FRB~121102 is an obvious case where birefringence might appear if bursts are detected at frequencies below 1~GHz. To date, bursts have been measured only at frequencies from 1.2 to 8 GHz. Typical parameters are peak flux densities of $\sim 0.1$-10~Jy and widths (FWHM) $W \sim 0.2$ to 8 ms \citep{Spitler+14, Spitler+16, Scholz+17, Michilli2018, Gajjar2018, Spitler2018, Hessels2018}. While Figure~\ref{fig:f3} suggests the significance of $\Delta \tg$ for modelling burst birefringence, the unknown values of $\lperp$ and $\DM_{\ell}$  for FRB 121102 leave $\Delta \tg$ largely unconstrained. Therefore, we only consider the TOA delay from $\Delta t_{\RM}$ for the following discussion. \\

Burst RMs  from FRB 121102 were initially  measured to be  $1.03 \times 10^5 \ \RMunits$  but  were $\sim 10\%$ lower half a year later \citep{Michilli2018} and have continued to decline (J. Hessels, private communication). In the following, we use $\RM = 10^5\ \RMunits$, close to the initial value measured by \citet{Michilli2018}, as a fiducial value. Owing to the finite redshift of the FRB 121102 source ($z = 0.193$, \citet{Tendulkar2017}), the observed RM appearing in \Eq \ref{eq:td}  is smaller than the source frame RM value ($\RM_{\text{src}}$) by a factor $(1+z)^2$. In terms of $\RM_{\text{src}}$ and the source frame frequency $\nu_{\text{src}} = \nu (1+z)$, $\Delta t_{\RM}$ is given by:
\be\label{eq:redshifted_tRM}
\Delta t_{\RM} = 57.2 \text{ ps} \left(\frac{\RM_{\text{src}}}{1 \  \RMunits} \right)\left( \frac{\nu_{\text{src}}}{1 \text{ GHz}} \right)^{-3}  (1+z).
\ee
\begin{figure*}[ht!]
\begin{center}
\includegraphics[width=0.9\linewidth]{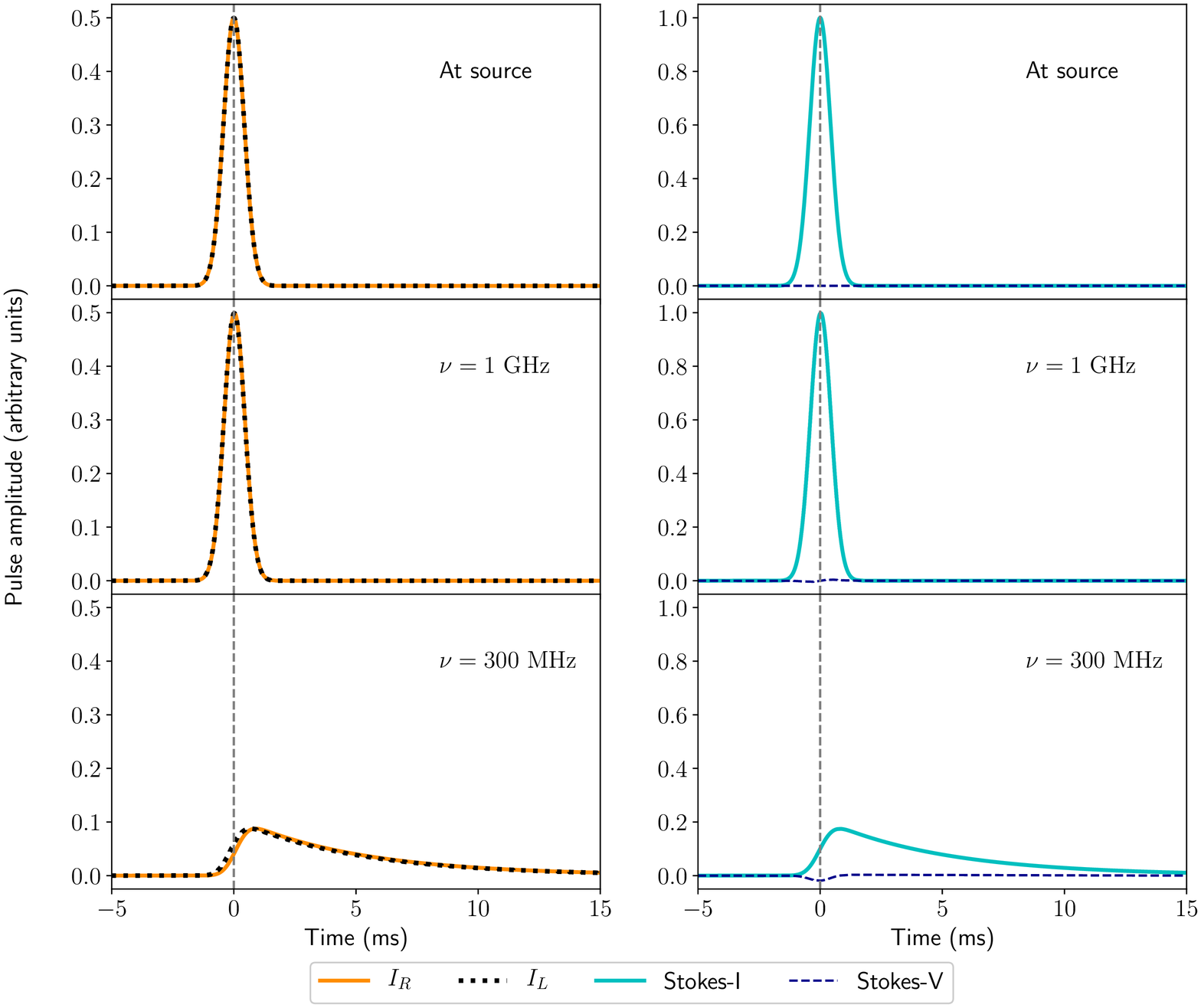}
\caption{Similar to Figure~\ref{fig:lrprofiles} except that pulse broadening from scattering has also been included as described in the text.   The top row depicts the pulse profiles at the source.  The second and third rows show the resulting profiles at $\nu = 1 \text{ GHz}$  and $\nu= 300 \text{ MHz} $.  Panels in the  left column show RHCP (orange solid line) and LHCP (black dotted line) profiles. The right column comprises of panels depicting pulse profiles in Stokes-I (cyan solid line) and Stokes-V (dark blue dashed line). The gray dashed line visible in all panels represents the time origin. 
\label{fig:lrscatt_profiles}}
\end{center}
\end{figure*}

To model the effects of birefringence and scattering on received bursts, we consider a Gaussian pulse with FWHM $W = 1 \text{ ms}$, i.e., width  $\sigma_t = W /  \sqrt{8\ln 2} = 0.43$~ms. 
We assume that the emitted pulse is 100\% linearly polarized, so the Stokes-I pulse can be expanded into a sum of equal-amplitude intensities for the two hands of  circular polarization. The initial pulse produced at the source is therefore: 
\be
I_{\rm R}(t) &=& I_{\rm L}(t) =  (1/2) e^{-t^2 / 2\sigma_t^2} ,
\\
I(t) &=& \IR(t) + \IL(t) = e^{-t^2/2\sigma_t^2} , 
\\
V(t) &=& \IR(t) -\IL(t) = 0.
\ee 
More generally, this treatment also applies to the case where the emitted radiation is unpolarized, which yields the same superposition of RHCP and LHCP pulses. 

Accounting for the relative delay $\Delta t_{\RM}$, Figure~\ref{fig:lrprofiles} shows the net pulse profiles in the RHCP, LHCP, Stokes-I and Stokes-V intensities at frequencies of 1 GHz, 300 MHz and 100 MHz. The  pulse profile at 1~GHz  is visually identical to the pulse profile at the source even with the large RM ($\approx 10^5 \ \text{rad m}^{-2}$) of FRB 121102. This is expected due to the fact that $\Delta t_{\RM} \approx 6 \ \mu \text{s} \ll \sigma_t$ at $\nu = 1 \text{ GHz}$.  However, at $\nu = 300 \text{ MHz}$ (third row), the differential arrival times of the  RCHP and LHCP pulses are clearly discernible. This gives rise to  a non-zero Stokes-V profile (right panel in the  third row that has the shape proportional to the derivative of the pulse profile in Stokes-I).  This signature in Stokes-V can serve as a useful diagnostic for identifying pulse splitting for initially linearly polarized or unpolarized waves. At the even lower radio frequency of 100~MHz, the temporal separation $\Delta t_{\RM} \approx 6 \text{ ms}$ between the RHCP and LHCP pulses exceeds $\sigma_t$. An observer then views the initial FRB pulse as a train of two pulses comprising of a LHCP pulse followed by a RHCP pulse, as depicted in the  left panel of the bottom row. For this case, the Stokes-I and Stokes-V profiles show complete separation of the two pulses resulting from the birefringence.  

\subsection{Effects of Scattering on Pulse Shapes}\label{sec:scattering}

\begin{figure*}[ht!]
\begin{center}
\includegraphics[width=0.9\linewidth]{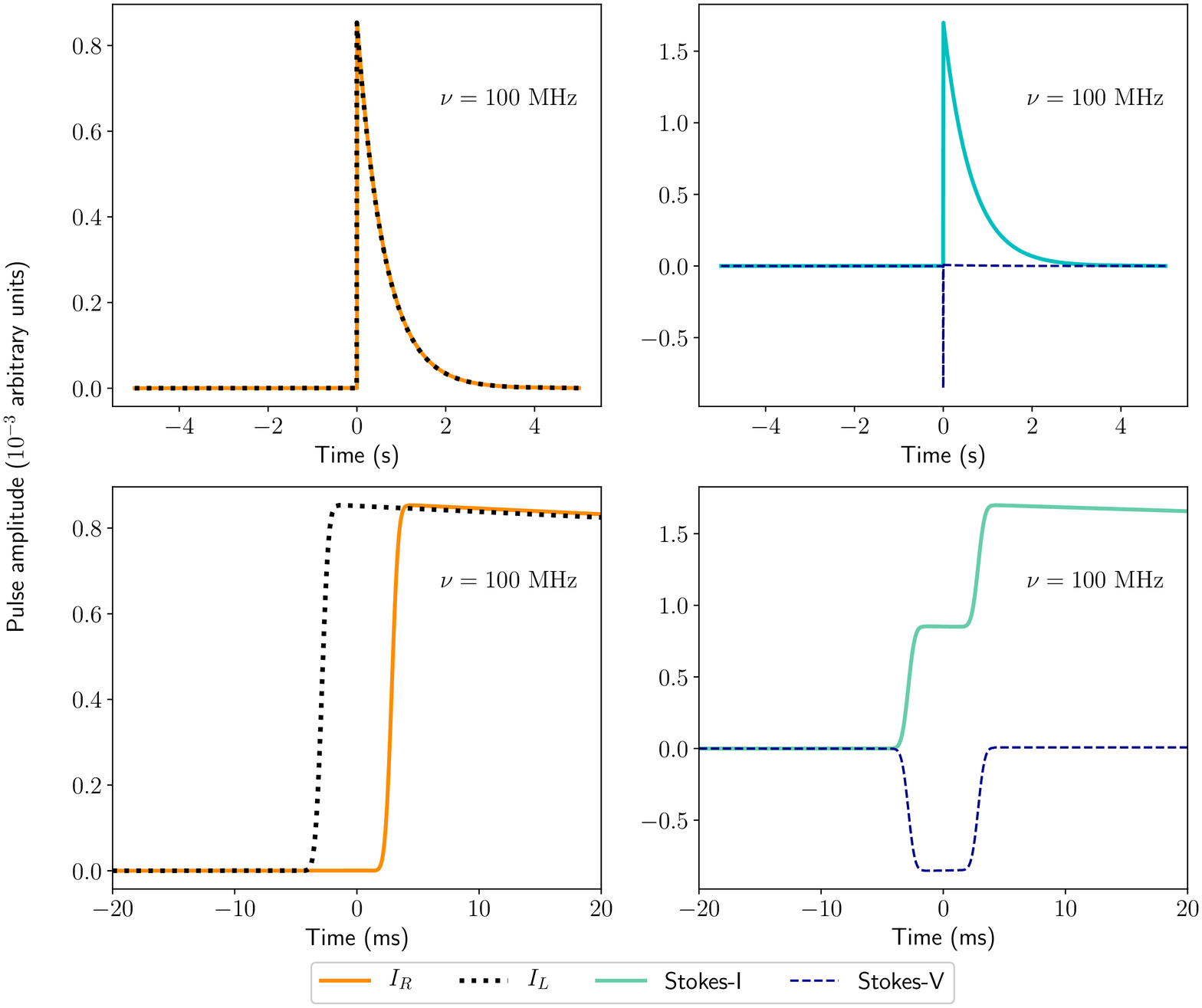}
\caption{Similar to Figure~\ref{fig:lrscatt_profiles} for $\nu= 100 \text{ MHz}$. Left column: Pulse profiles in RHCP (orange solid line) and LHCP (black dotted line). Right column: Pulse profiles in Stokes-I (cyan solid line) and Stokes-V (dark blue dashed line). The  amplitudes on the y-axis of all panels are a factor of $10^{3}$ smaller than those shown in Figures~\ref{fig:lrprofiles} and \ref{fig:lrscatt_profiles} owing to the large scattering time that broadens the pulse over  a long time, thus, reducing its amplitude. The time axis in the top panel extends over $\pm 5$~s, much longer than the 20~ms extent in Figure~\ref{fig:lrscatt_profiles}. The bottom panels zoom-in near the time origin to show pulse splitting. 
\label{fig:lrscatt100}}
\end{center}
\end{figure*}

Measured pulses are the convolution of intrinsic shapes with a pulse broadening function (PBF)  $p(t)$  that is asymmetric and causal.    The PBF for a  medium with Kolmogorov electron density fluctuations rises sharply and decays more slowly than an exponential \citep{Lambert1999}; however, for our purposes, an exponential form 
$p(t) = \taud^{-1} e^{-t/\taud}H(t)$, where $H(t)$ is the Heaviside step function and $\taud$ is the pulse broadening time, suffices to demonstrate the interplay of scattering and birefringence. Scatter broadening can distort burst structure and hide the effects of birefringence.

We evaluate  $\taud$ for the FRB~121102 by using the scintillation bandwidth $\dnud$ and the `uncertainty' relation 
\be
\taud = \frac{C_1}{2\pi \Delta \nu_d} 
\label{eq:taud}
\ee
where  $C_1 \sim 1$ \citep{Lambert1999}. Through detailed analysis of the time-frequency structure of the bursts seen in FRB 121102 observations, \citet{Michilli2018}, \citet{Gajjar2018}, \citet{Spitler2018} and \citet[][in preparation]{Hessels2018}  have independently estimated the burst scintillation bandwidths at different frequencies. \citet{Hessels2018} report $\dnud = 58.1 \text{ kHz}$ at $\nu = 1.66 \text{ GHz}$, a value that is consistent with the high-frequency values using a scaling law $\dnud \propto \nu^{-\alpha}$ with $\alpha = 4.4$. This yields a pulse broadening time $\taud \sim 25.4 \, \mu s\, \nu^{-4.4}$ with $\nu$ in GHz. 

Taking scattering and birefringence into consideration, Figure~\ref{fig:lrscatt_profiles} shows pulse profiles at 1~GHz  (middle row) and 300~MHz (bottom row), where (from Equation~\ref{eq:taud}) $\taud$ equals $25.4 \ \mu \text{s}$ and 5 ms,  respectively. The small value of $\taud$ relative to $\sigma_t$ at $\nu = 1 \text{ GHz}$ results in the absence of a significant signature of scattering at this frequency. At $\nu = 300 \text{ MHz}$, $\taud$ is comparable to $\sigma_t$, and scattering smears out the RHCP and LHCP pulses towards $t>0 \text{ ms}$. As illustrated in the bottom row of Figure~\ref{fig:lrscatt_profiles}, the pulses are asymmetric with  exponential tails toward later arrival times. However, despite the visible effects of scattering, pulse splitting still gives rise to a temporal lag of the RHCP pulse with respect to the LHCP pulse. This is manifested as a dip in the Stokes-V profile (right panel in bottom row of Figure~\ref{fig:lrscatt_profiles}) around $t=0$.   The dip becomes more prominent and long-lived at low radio frequencies where  pulse splitting effects are stronger. Thus, at $\nu = 100 \text{ MHz}$, a conspicuous dip in the Stokes-V profile can be seen despite the dominant effects of scattering on the overall pulse profile in $I_R$, $I_L$ and Stokes-I. This is illustrated in the panels constituting Figure~\ref{fig:lrscatt100}.  Equation~\ref{eq:taud} yields a  pulse broadening time $\taud \sim 0.62$~s at $\nu = 100 \text{ MHz}$, which causes the exponential scattering tails of the RHCP and LHCP pulses to extend out to times as large as $t=2\text{ s}$, much larger than either the intrinsic pulse width or the birefringence splitting time.

\section{The Galactic Center Magnetar (J1745$-$2900)}\label{sec:Galcenter_magnetar}

The Galactic center magnetar has been known to emit linearly polarized pulses (fractional linear polarization $\ge 80\%$) between $4.5-20 \text{ GHz}$ \citep{Shannon2013}. Its large RM ($\approx - 6.6 \times 10^4 \ \RMunits$) then makes it a suitable candidate for observing pulse birefringence at low radio frequencies. As its RM is similar in magnitude to that of FRB 121102, $\Delta t_{\RM}$ is expected to be as shown in Figure~\ref{fig:lrprofiles}. However, pulse splitting is reversed in polarization due to the negative sign of its RM. \citet{Spitler2014magnetar} measured a pulse broadening time scale, $\tau_d \sim 1.3~s \ \nu^{-3.8}$ ($\nu$ in GHz) for pulses from the Galactic center magnetar at a frequency of $1 \text{ GHz}$. The consequent large pulse smearing from scattering at low radio frequencies can significantly lower pulse peak amplitudes and thereby, diminish the detectability of any effects of $\Delta t_{\RM}$ on the Stokes-V profile (dip seen in lower right panel of Figure~\ref{fig:lrscatt100}, which shall manifest as a peak for pulses from the Galactic center magnetar owing to its negative RM). 

While we expect scattering to mask $\Delta t_{\RM}$, Figure~\ref{fig:f3} suggests an absolute geometric splitting time, $|\Delta \tg| \approx ~ 1.8~s$ at $\nu =800 \text{ MHz}$ for $\lperp = 1 \text{ AU}$, which is comparable to the pulse broadening time ($\tau_d\approx 3~s$ at $\nu = 800 \text{ MHz}$) and the rotational period ($P = 3.76~s$ \citep{Mori2013}) of the magnetar. Unlike birefringence from $\Delta t_{\RM}$, pulse splitting from $\Delta \tg$ should therefore produce visible fluctuations in Stokes-V for the Galactic center magnetar. 

We note that our discussions of $\Delta \tg$ and $\tau_d$ in \S~\ref{sec:refraction} and \S~\ref{sec:scattering}, respectively, assume refraction and scattering from a thin screen in the ISM intervening our line of sight to the source. To physically interpret the large RM variations ($\sim 3500\ \RMunits$ in 4 years) observed for the Galactic center magnetar, \citet{Desvignes2018} invoke a two-screen model comprising of a nearby thin ($\sim 0.1 \text{ pc}$ from the magnetar) screen that accounts for magnetic field fluctuations, and a distant ($\sim 6 \text{ kpc}$ from the magnetar) second screen that explains temporal pulse broadening. Multi-path propagation from multiple screens could lead to estimates of $\Delta \tg$ and $\tau_d$ which substantially differ from that discussed above. 

\section{Nanoshots from the Crab Pulsar}\label{sec:Crab}

\citet{Hankins2003} and \citet{Hankins2007} identified very narrow ($W_{\Delta} \sim $ ns) shot pulses as structure within giant pulses (GPs) detected at the rotational phase of the inter-pulse emission from the Crab pulsar at $\nu \simeq 8~\text{GHz}$. In some of these inter-pulse GPs, the set of nanoshots comprising a GP emission envelope shows both hands of circular polarization with seemingly stochastic flipping of the sign of Stokes-V between consecutive nanoshots \citep{Jessner2010,Hankins2016}. In others, there is nearly 100\% linear polarization with very little circular polarization. These results suggest that the GP and nanoshot properties are episodic, perhaps arising from propagation effects.   Given the very narrow durations, nanoshots are promising candidates for showing birefringence.   

The average, large-scale RM across the Crab Nebula is $\RM \sim -21$~rad~m$^{-2}$  \citep[][]{1991ApJ...368..231B}. Assuming $\RM_{\text{f}} = -21 \ \RMunits$, Figure \ref{fig:f3} informs us that $\Delta t_g < \Delta t_{\RM} \sim 10 - 100 \text{ ps}$  at $\nu = 5-10$ GHz. However, radio depolarization indicates that small filaments  in the nebula with scale sizes $\sim 1000$~AU have a filling factor $\sim 30$\% and internal RM values with a standard deviation $\sim 400$~rad~m$^{-2}$ \citep[][]{1991ApJ...368..231B}. Radio wave propagation through filaments with such RM values could lead to $\Delta t_{\RM} \sim 1-10$ ns  at $\nu = 5-10$ GHz. Thus, nanosecond duration pulse splitting from $\Delta t_{\RM}$ combined with changes in geometry from motion of the Crab pulsar could potentially produce episodic birefringent effects that accompany plasma lensing events reported by \citet{bwv00} and \citet{lpg01}.

We consider an individual nanoshot that is linearly polarized when emitted. The electric field associated with this nanoshot can be described by $\varepsilon_{\rm \ell}(t) = \varepsilon(t) (\cos2\psi\xhat + \sin2\psi\yhat)$, where
$\psi$ is the polarization angle. In terms of the units vectors $\epspm$, the circularly polarized field components are a sequence of shot pulses with shape $\Delta(t)$ and amplitude $a_j$.
\be
\varepsilon_{\pm}(t) = \frac{1}{\sqrt{2}} \sum_j a_j e^{\pm 2i\psi}\Delta(t-t_j)
. 
\ee
After propagation that may impart different RM and DM values to different nanoshots if there is multipath propagation, these become
\be
\varepsilon^{\rm (p)}_{\pm}(t) = \frac{1}{\sqrt{2}} 
\sum_j a_j e^{\pm 2i\psi}
\Delta(t-t_j - t_{\rm DM_j} \pm t_{\rm RM_j} ) 
. 
\ee
 
On nanoshot time scales, the polarization signature in the Stokes parameters  depends on the degree of overlap between the split RHCP and LHCP components.  
If $\Delta t_{\rm RM} \ll W_{\Delta}$,
the nanoshot (again assumed linearly polarized when emitted) will be linearly polarized and show Faraday rotation.  In the 
opposite case with $\Delta t_{\rm RM} \gg W_{\Delta}$, the lack of overlap implies that the two separate nanoshots will have opposite senses of CP.   With partial overlap, the time sequence is first one hand of CP followed by an interval of linear polarization followed by the other hand of CP.   At any instant, the propagated nanoshot is 100\% polarized.  

For an intrinsically unresolved shot pulse whose width is determined by the receiver bandwidth $\Delta\nu$,
$W_\Delta = 1 / \Delta\nu = 1$~ns for a 1~GHz bandwidth.   
Birefringent splitting yields overlapped CP components if
$ \vert \RM \vert \le 17.5\  {\rm rad \ m^{-2}}\ \nu^3 / \Delta\nu$  for $\nu$ and $\Delta\nu$ in GHz.  

\section{Two Dimensional Coherent  Dedispersion}
\label{sec:deDMRM}
To date, burst detection algorithms implemented as part of FRB surveys mostly employ the incoherent dedispersion technique that introduces instrumental broadening of bursts. The best resolution of the intrinsic burst width is obtained with coherent dedispersion which unwraps the frequency-dependent phase imposed by propagation through intervening plasma \citep[][]{hr75}.  In simple terms, this involves multiplying the Fourier components of voltage data obtained from each polarization channel by the inverse function, 
$e^{-i\phi(\nu)}$, where
$\phi(\nu) = \int_0^d ds\, k(\nu)$.
Conventionally, the phase involves only the \DM,  $\phi(\nu, \DM) = -(c r_e/\nu)\DM$, and dedispersion is applied to sampled baseband data over a bandwidth originally centered on a center frequency $\nu_0$. 

Dedispersion is extended to two parameters, \RM\ as well as \DM, by using the phase in Equation~\ref{eq:phi1}-\ref{eq:phi2}, which we rewrite here as
\be
\phi(\nu, \DM, \RM) = 
-(c r_e/\nu)\DM \mp (c/\nu)^2 \RM.  
\label{eq:phiDMRM}
\ee
Expanding about $\nu_0$ using $\nu = \nu_0 + \delta\nu$,  there are contributions from both \DM\ and \RM\ to terms that are  linear, quadratic, and cubic, etc. in $\delta\nu/\nu_0$, where the linear term affects the arrival time but not the shape of the dedispersed pulse. 

Two-parameter dedispersion can be applied in several ways to identify the  birefringence.   The total intensity (Stokes $I$) can be maximized to identify `best' estimates, $\DMhat$ and $\RMhat$. This approach would align shot pulses that would otherwise display the \RM\ splitting time.  Presumably, intensity maximization would also minimize the overall  width of a pulse comprising multiple nanoshots. 
However, it is conceivable that, due to multipath propagation,  different nanoshots might yield different values for $\DMhat$ and $\RMhat$. An alternative approach would be to minimize TOA differences between nanoshot features in the separate CP channels. 

Other Stokes parameters can be investigated in addition to Stokes-I. If intrinsic linear polarization is assumed,  Stokes-V could be minimized in order to identify $\DMhat$ and $\RMhat$.  If, however, the intrinsic polarization is circular, there would be no preferred value for $\RM$. 

Finally, one expects the dispersion law near a pulsar (or FRB source) to differ from that of a cold, weakly magnetized plasma, so departures from Equation~\ref{eq:phiDMRM} may be manifested as an inconsistency between the quadratic and cubic terms of the expanded phase. 

\section{Noise Statistics} \label{sec:noise}
Pulsar and FRB pulse shapes are consistent with a statistical model where the electric field is a pulse envelope that modulates a zero-mean noise process. Physically, the noise process is made up of coherent shot pulses with widths comparable to or narrower than the inverse radio frequency, $\nu^{-1}$.  If shot pulses are closely spaced, the noise will have Gaussian statistics
\citep[amplitude modulated noise (AMN),][]{1975ApJ...197..185R}. But, a sparse sequence like the nanoshots of the Crab pulsar will show non-Gaussian statistics
\citep[amplitude modulated shot noise (AMSN),][]{1976ApJ...210..780C}.  Alternatively, the envelope could correspond to a change in the rate of shot pulses
\citep[rate modulated shot noise (RMSN),][]
{cordes_wasserman16}.

The emitted noise process determines the measured degree of polarization, typically made with time resolution $\gg \nu^{-1}$, if propagation does not alter the polarization state and if Faraday rotation does not cause
any bandwidth depolarization.  However, strong birefringent pulse splitting can convert emitted linear polarization into circular polarization. This can also induce circular polarization from an unpolarized noise process, as discussed previously. An emitted linearly polarized pulse may be distinguished from an intrinsically unpolarized pulse through noise statistics using high time resolution data.

If the intrinsic polarization is 100\% linearly polarized, the intensities measured in RHCP and LHCP will have identical envelopes and noise\footnote{
For the sake of presentation, we assume that receiver noise is negligible.}, but will be offset in time due to birefringence.  

The cross correlation function (CCF) of the RHCP and LHCP  intensities,
$C_{\rm RL}(\tau) = 
 \left\langle  \IR(t) \IL(t+\tau) \right\rangle$,
 will then show a peak at a non-zero time lag equal to the birefringent splitting time.  
Letting the linearly polarized field be $\varepsilon_{\rm \ell}(t) = \varepsilon(t) (\cos2\psi\xhat + \sin2\psi\yhat)$, the propagated R and L fields are 
\be
\varepsilon_{\rm R,L}(t) = 
	\frac{1}{\sqrt{2}}
     	{\varepsilon(t-t_{\pm})e^{\pm 2i \psi}} 
    \epspm.
\ee
The intensities then differ only in their relative delays. So, the intensity CCF is
\be
C_{\rm RL}(\tau) = R_I(\tau + t_+ - t_-),
\ee
where $R_{I}(\tau)$ is the autocorrelation function (ACF) of $I(t) = \vert \varepsilon(t) \vert^2$.
Evaluating for an AMSN model, 
$\varepsilon(t) = a(t) m(t)$, where $a(t)$ is the deterministic real amplitude and $m(t)$ is complex noise with time stationary statistics. We define $A = a^2$ and  $M = \vert m \vert^2$, and  let $M$ have unit mean, $\left\langle M \right\rangle = 1$.
The ACF of $M$ becomes
\be
\left\langle M(t) M(t+\tau) \right\rangle
	= 
    1 + \mu^2 r(\tau),
\ee
where $\mu = \left( \left\langle M^2 \right\rangle - 1 \right)^{1/2}$ is the modulation index and $r(\tau)$ is a narrow function whose  width is equal to the time resolution determined by a receiver bandwidth (ranging from sub-ns to 1~$\mu s$).  Complex Gaussian noise statistics correspond to $\mu = 1$. 
Then, 
\be
C_{\rm RL}(\tau) &=& 
	A(t-t_+) A(t+\tau- t_- )
    \left[ 1 + \mu^2 \Delta(\tau + t_+ - t_-) \right]
    \nonumber \\
    &\approx& A(t) A(t+\tau) \left[ 1 + \mu^2 r(\tau + t_+ - t_-)\right],
\ee
where the approximate equality applies if
the splitting angle is much smaller than the width of $A$. 

On the other hand, if an emitted burst is  unpolarized, the noise processes in RHCP and LHCP will be uncorrelated and the cross correlation will vanish.

\section{Summary and Conclusions}\label{sec:summary}

In this paper, we have considered observable manifestations of radio wave propagation through magnetized plasmas besides the standard Faraday rotation of the polarization ellipse. Arrival time variations arising from birefringence are typically too small to measure directly for most pulsars. But, in extreme conditions where RMs are very large or pulses are extremely narrow, TOA effects can become measurable. 

We have identified  two distinct contributions to pulse TOAs, one due to the difference in group velocity of RHCP and LHCP pulses and the other due to differential refraction. For FRB 121102, the Galactic center magnetar, and the Crab pulsar, we have quantified the spectral regimes where refraction dominates group-velocity effects and vice versa. 
 
Birefringent TOAs may be relevant to fast radio bursts observed at low frequencies. The repeating FRB121102 has a high enough RM that polarization splitting at $100-300$ MHz can be larger than (intrinsic) burst widths, though multi-path scattering likely will broaden bursts significantly at those frequencies. 

Splitting is in principle relevant to the high-RM line of sight to the Galactic-center magnetar J1745$-$2900, though intense scattering will prevent measurement of splitting times at sub-GHz frequencies.   

The Crab pulsar, however, shows nanoshot pulses that occasionally display both RHCP and LHCP that may arise from splitting due to propagation through dense filaments in the Crab Nebula. To aid further study of this possibility, we have outlined a two-parameter, coherent dedispersion method that we will explore further. 

\begin{acknowledgements}
We thank Shami Chatterjee and Tim Hankins for useful conversations and correspondence.  The authors acknowledge support from  the NANOGrav Physics Frontier Center (NSF award 1430284). 
\end{acknowledgements}

\bibliography{aobook,books,magnetoionic,psrrefs,NASA_FRB.bib,noise.bib}
\end{document}